\documentstyle[12pt,amssym]{article}

\newcommand{\ab}{\bar{a}}
\newcommand{\cb}{\bar{c}}
\newcommand{\db}{\bar{d}}
\newcommand{\lb}{\bar{l}}
\newcommand{\mb}{\bar{m}}
\newcommand{\lambdab}{\bar{\lambda}}
\newcommand{\psib}{\bar{\psi}}
\newcommand{\yyb}{\bar{Y}}
\newcommand{\hhb}{\bar{H}}
\newcommand{\llb}{\bar{L}}
\newcommand{\hht}{\tilde{H}}
\newcommand{\bh}{\hat{b}}
\newcommand{\ch}{\hat{c}}
\newcommand{\lh}{\hat{l}}
\newcommand{\mh}{\hat{m}}
\newcommand{\lambdah}{\hat{\lambda}}
\newcommand{\psih}{\hat{\psi}}
\newcommand{\yyh}{\hat{Y}}
\newcommand{\hhh}{\hat{H}}
\newcommand{\llh}{\hat{L}}
\newcommand{\hhc}{\check{H}}
\def\R{\mbox{$\Bbb R$}}
\def\case#1#2{{\textstyle{#1\over #2}}}

\title{
Connection between type B (or C) and F factorizations and construction of algebras}
\author{A. Del Sol Mesa $^{a,}$\thanks{E-mail address: A.DelSolMesa@cs.ucl.ac.uk}\
,  C. Quesne $^{b,}$\thanks{Directeur de recherches FNRS; E-mail address:
cquesne@ulb.ac.be}\\ 
$^a$ {\small Bioinformatics Unit, Department of Computer Science, University
College London,}\\ 
{\small Gower Street, London WC1E 6BT, UK}\\ 
$^b$ {\small Physique Nucl\'eaire Th\'eorique et Physique
Math\'ematique,  Universit\'e Libre de Bruxelles,} \\ 
{\small Campus de la Plaine CP229, Boulevard~du Triomphe, B-1050 Brussels,
Belgium}}
\date{ }
\begin{document}
\baselineskip=20pt plus 1pt minus 1pt
\maketitle
%
%
\newpage
\noindent
{\bf Abstract.} 
In a recent paper (Del Sol Mesa A and Quesne C 2000 {\em J.\ Phys.\ A: Math.\ Gen.}
{\bf 33} 4059), we started a systematic study of the connections among different
factorization types, suggested by Infeld and Hull, and of their consequences for the
construction of algebras. We devised a general procedure for constructing satellite
algebras for all the Hamiltonians admitting a type E factorization by using the relationship
between type A and E factorizations. Here we complete our analysis by showing that for
Hamiltonians admitting a type F factorization, a similar method, starting from either type B
or type C ones, leads to other types of algebras. We therefore conclude that the existence
of satellite algebras is a characteristic property of type E factorizable Hamiltonians. Our
results are illustrated with the detailed discussion of the Coulomb problem.
\par
%
%
\newpage
\section{Introduction}

In a recent paper~\cite{delsol00} (henceforth referred to as I and whose equations
will be quoted by their number preceded by I), we investigated the role of the
factorization method~\cite{infeld, carinena} in the construction of a new class of
symmetry algebras, called {\sl satellite algebras}, introduced in a previous
work~\cite{delsol98}. Such algebras, generalizing the so-called potential
algebras~\cite{alhassid, frank}, depend upon some auxiliary variables and connect
among themselves wavefunctions belonging to different satellite potentials and
corresponding to different energy eigenvalues (instead of the same ones). We also
devised a general procedure for determining an $\mbox{\rm so(2,2)} \simeq 
\mbox{\rm su(1,1)} \oplus \mbox{\rm su(1,1)}$ or $\mbox{\rm so(2,1)} \simeq
\mbox{\rm su(1,1)}$ satellite algebra for all the Hamiltonians admitting a type E
factorization by using the known relationship between type A and E factorizations
and an algebraization similar to that used in the construction of potential
algebras.\par
%
%
The purpose of the present work is to apply a similar procedure to the
Hamiltonians admitting a type F factorization by starting this time from type B or
C one. Let us recall that in such cases, the functions $r(x,m)$, $k(x,m)$,
and $L(m)$ entering the factorization method formalism (see section~2 of~I for a
summary of the latter) are given in terms of some constants $a$, $b$, $c$, $d$, and
$q$ by
\begin{eqnarray}
  \mbox{\rm Type B}: && r(x,m) = - d^2 e^{2ax} + 2ad \left(m + c +
           \frac{1}{2}\right) e^{ax} \\
  && k(x,m) = d e^{ax} - (m + c) a  \label{eq:B-k} \\
  && L(m) = - a^2 (m + c)^2  \label{eq:B-L} \\
  \mbox{\rm Type C}: && r(x,m) = - \frac{(m+c) (m+c+1)}{x^2} - \frac{1}{4} b^2 x^2 +
           b (m-c) \\
  && k(x,m) = \frac{m+c}{x} + \frac{1}{2} bx \\
  && L(m) = - 2bm + \frac{1}{2} b \\
  \mbox{\rm Type F}: && r(x,m) = - \frac{2q}{x} - \frac{m(m+1)}{x^2} \label{eq:F-r}
           \\
  && k(x,m) = \frac{m}{x} + \frac{q}{m} \label{eq:F-k} \\
  && L(m) = - \frac{q^2}{m^2} \label{eq:F-L} 
\end{eqnarray} 
respectively.\par
%
%
Since the precise relationship between type F and type B or C factorizations is not
detailed in~\cite{infeld}, we first derive it and then construct the corresponding
algebra in sections~2 and~3. The general results are illustrated on the example of
the Coulomb problem in section~4. Section~5 contains the conclusion.\par
%
%
\section{Case of type B and F factorizations}

\setcounter{equation}{0}
Let us consider the most general second-order differential equation admitting a
type F factorization. From equations~(I2.1) and~(\ref{eq:F-r}), it is given by
\begin{equation}
  \left(\frac{d^2}{dr^2} - \frac{2q}{r} - \frac{m(m+1)}{r^2} + \lambda\right) \psi(r)
  = 0 \label{eq:F-eq}
\end{equation}
where the variable $x$ is changed into $r$ and $\psi = Y_l^m$ denote the
normalized eigenfunctions corresponding to the discrete eigenvalues $\lambda =
\lambda_l$. Since we want to apply the theory to Hermitian Hamiltonians with
bound states, we restrict ourselves to the case where $q$ is some negative real
constant. Then $L(m)$, given in~(\ref{eq:F-L}), is an increasing function of~$m$ so
that the problem is of class~I, $m=0$, 1, \ldots,~$l$, and
\begin{equation}
  \lambda = L(l+1) = - \frac{q^2}{(l+1)^2}. \label{eq:F-lambda}
\end{equation}
In the corresponding $k(x,m)$, given in~(\ref{eq:F-k}), $m$ occurs in the
denominator, hence a straightforward algebraization of the ladder operators is
impossible.\par
%
%
To carry out such an algebraization, one may transform the type F factorizable
equation~(\ref{eq:F-eq}) into a type B or C one. In the present section, we consider
the former alternative, given by
\begin{equation}
  \left[\frac{d^2}{dx^2} - \db^2 e^{2\ab x} + 2 \ab \db \left(\mb + \cb + \frac{1}{2}
  \right) e^{\ab x} + \lambdab\right] \psib(x) = 0 \label{eq:B-eq} 
\end{equation}
where a bar is put on top of all the constants to distinguish them from those used
for type F factorization and the normalized eigenfunctions $\yyb_{\lb}^{\mb}$,
corresponding to the eigenvalues $\lambdab = \lambdab_{\lb}$, are denoted by
$\psib$. From equations~(I2.4), (\ref{eq:B-k}), and~(\ref{eq:B-L}), it follows that
the associated ladder operators $\hhb^{\pm}(\mb)$, which depend linearly on
$\mb$, and the real constant $\llb(\mb)$ can be written as 
\begin{eqnarray}
  \hhb^{\pm}(\mb) & = & \pm \frac{d}{dx} + \db e^{\ab x} - (\mb + \cb) \ab
       \label{eq:B-ladder} \\
  \llb(\mb) & = & - \ab^2 (\mb + \cb)^2. \label{eq:B-L-bis}
\end{eqnarray}
\par
%
%
By performing a change of variable and of function
\begin{equation}
  r = e^{\ab x} \qquad \psi(r) = e^{\ab x/2} \psib(x) \label{eq:F-B-change}
\end{equation}
equation~(\ref{eq:F-eq}) is transformed into an equation of type~(\ref{eq:B-eq}),
\begin{equation}
  \left[\frac{d^2}{dx^2} + \lambda \ab^2 e^{2\ab x} - 2q \ab^2 e^{\ab x} - \left(m +
  \frac{1}{2}\right)^2 \ab^2\right] \psib(x) = 0. \label{eq:F-B-eq} 
\end{equation}
Comparison between (\ref{eq:B-eq}) and~(\ref{eq:F-B-eq}) shows that the
parameters and eigenvalues of both factorization types are connected through the
relations
\begin{eqnarray}
  \frac{\db^2}{\ab^2} & = & - \lambda \label{eq:F-B-rel1} \\
  \frac{\db}{\ab} \left(\mb + \cb + \frac{1}{2}\right) & = & -q \label{eq:F-B-rel2} \\
  \frac{\lambdab}{\ab^2} & = & - \left(m + \frac{1}{2}\right)^2. \label{eq:F-B-rel3}
\end{eqnarray}
\par
%
%
{}From (\ref{eq:F-B-change}), it is clear that only real values of $\ab$ are to be
considered here. Hence from~(\ref{eq:B-L-bis}), it results that the type B problem
is of class~II, $\mb = \lb$, $\lb+1$,~\ldots, and
\begin{equation}
  \lambdab = \llb(\lb) = - \ab^2 (\lb + \cb)^2. \label{eq:B-lambda}
\end{equation}
Equation~(\ref{eq:F-B-rel2}) and the assumption $-q \in \R^+$ imply that $\mb +
\cb + \frac{1}{2}$ and $\db/\ab$ both belong to $\R^+$ or to $\R^-$. The latter
alternative is excluded because $\mb$ may take arbitrary large positive
values. Moreover we may always assume that $\ab \in \R^+$ since changing the sign
of $\ab$ can be compensated by the change of variable $x \to -x$. We therefore
conclude that the parameters of type B and F factorizations may be restricted to
values such that $\ab$, $\db$, $\mb + \cb + \frac{1}{2}$, and $-q \in \R^+$.\par
%
%
In such a case, equations (\ref{eq:F-B-rel1}) -- (\ref{eq:F-B-rel3}), together
with~(\ref{eq:F-lambda}) and~(\ref{eq:B-lambda}), lead to the relations
\begin{equation}
  \frac{\db}{\ab} = - \frac{q}{l+1} = \sqrt{-\lambda} \qquad \mb + \cb = l + 
  \frac{1}{2} \qquad \lb + \cb = m + \frac{1}{2}. \label{eq:F-B-solrel}
\end{equation}
\par
%
%
On using (\ref{eq:F-B-change}) and~(\ref{eq:F-B-solrel}), the type B ladder
operators~(\ref{eq:B-ladder}) lead to ladder operators for the original type F
eigenfunctions $\psi$,
\begin{eqnarray}
  \hht^{\pm}(l) & \equiv & e^{\ab x/2} \hhb^{\pm}(\mb) e^{-\ab x/2} \nonumber \\
  & = & \ab \left[\pm r \frac{d}{dr} + \sqrt{-\lambda}\, r - \left(l + \frac{1}{2} \pm
        \frac{1}{2}\right)\right].
\end{eqnarray}
\par
%
%
{}From these ladder operators, we can get Lie algebra generators by introducing an
auxiliary variable $\eta \in [0, 2\pi)$ and extended eigenfunctions defined by
\begin{equation}
  \Psi_t(r, \eta) = (2\pi)^{-1/2} e^{{\rm i} t \eta} \psi(r) \label{eq:F-B-extended}
\end{equation}
where
\begin{equation}
  t \equiv l+1. \label{eq:t}
\end{equation}
Since 
\begin{equation}
  T_0 = - {\rm i} \frac{\partial}{\partial \eta} \label{eq:T0}
\end{equation}
is such that
\begin{equation}
  T_0 \Psi_t(r, \eta) = t \Psi_t(r, \eta) \label{eq:T0-action}
\end{equation}
we may replace t by $- {\rm i} \partial/\partial \eta$ when dealing with extended
eigenfunctions. By combining the transformation
\begin{equation}
  \ab^{-1} e^{\pm {\rm i} \eta} \hht^{\mp}\left(l + \case{1}{2} \pm \case{1}{2}
  \right) \to T_{\pm}
\end{equation}
with this substitution, we obtain
\begin{equation}
  T_{\pm} = e^{\pm {\rm i} \eta} \left(\mp r \frac{\partial}{\partial r} + {\rm i}
  \frac{\partial}{\partial \eta} + \sqrt{-\lambda}\, r\right). \label{eq:Tpm}
\end{equation}
\par
%
%
It is straigthforward to check that the operators $T_0$, $T_+$, $T_-$ close an 
$\mbox{\rm su(1,1)} \simeq \mbox{\rm so(2,1)}$ Lie algebra, i.e.,
\begin{equation}
  [T_0, T_{\pm}] = \pm T_{\pm} \qquad [T_+, T_-] = - 2T_0
\end{equation}
and that the Casimir operator of the latter can be written as
\begin{eqnarray}
  C & \equiv & - T_+ T_- + T_0 (T_0 - 1) \nonumber \\
  & = & r^2 \left(\frac{\partial^2}{\partial r^2} - 2 {\rm i} \frac{\sqrt{-\lambda}}{r}
       \frac{\partial}{\partial \eta} + \lambda\right).
\end{eqnarray}
\par
%
%
{}From (\ref{eq:F-eq}), (\ref{eq:F-lambda}), (\ref{eq:t}), (\ref{eq:T0}),
and~(\ref{eq:T0-action}), it results that the action of $C$ on the extended
eigenfunctions~(\ref{eq:F-B-extended}) is given by
\begin{equation}
  C \Psi_t(r, \eta) = m(m+1) \Psi_t(r,\eta).
\end{equation}
The (nonunitary) su(1,1) irreducible representations spanned by $\Psi_t(r,\eta)$
may therefore be characterized by $m$ and such functions may be denoted by
$\Psi^{(m)}_t(r, \eta)$. It should be stressed that the su(1,1) generators $T_{\pm}$
change $t$ into $t \pm 1$, while leaving $m$ and $\lambda$ unchanged. From the
first relation in~(\ref{eq:F-B-solrel}), it is clear that $q$ becomes $q' = q (l+1\pm
1)/(l+1) = q (t\pm1)/t$. Hence $T_{\pm}$ connect among themselves eigenfunctions
corresponding to different potentials but to the same energy eigenvalue. We
therefore conclude that the su(1,1) algebra resulting from type B factorization is a
potential algebra.\par
%
%
\section{Case of type C and F factorizations}

\setcounter{equation}{0}
Let us now turn ourselves to the latter alternative, i.e., transforming the type F
factorizable equation~(\ref{eq:F-eq}) into a type C one. For such a purpose, we shall
take advantage of the relation between types F and B already established in
section~2 and of the equivalence between types B and C noted in~\cite{infeld}.\par
%
%
More specifically, the type B factorizable equation~(\ref{eq:B-eq}) can be
transformed into a type C one,
\begin{equation}
  \left[\frac{d^2}{dy^2} - \frac{(\mh+\ch) (\mh+\ch+1)}{y^2} - \frac{1}{4} \bh^2 y^2
  + \bh (\mh-\ch) + \lambdah\right] \psih(y) = 0 \label{eq:C-eq}
\end{equation}
where the normalized eigenfunctions $\yyh_{\lh}^{\mh}$ corresponding to the
eigenvalues $\lambdah = \lambdah_{\lh}$ are denoted by $\psih$, by the change of
variable and of function
\begin{equation}
  x = \frac{2}{\ab} \ln \frac{y}{2} \qquad \psib(x) = y^{-1/2} \psih(y).
  \label{eq:B-C-change}
\end{equation}
We indeed obtain
\begin{equation}
  \left[\frac{d^2}{dy^2} + \frac{(4 \lambdab/\ab^2) + (1/4)}{y^2} -
  \frac{\db^2}{4\ab^2} y^2 + \frac{2\db}{\ab} \left(\mb + \cb +
  \frac{1}{2}\right)\right] \psih(y) = 0
\end{equation}
which coincides with~(\ref{eq:C-eq}) provided
\begin{eqnarray}
  (\mh+\ch) (\mh+\ch+1) & = & - \frac{4\lambdab}{\ab^2} - \frac{1}{4} 
       \label{eq:B-C-rel1} \\
  \bh^2 & = & \frac{\db^2}{\ab^2} \\
  \bh (\mh-\ch) + \lambdah & = & \frac{2\db}{\ab} \left(\mb+\cb+
       \frac{1}{2}\right). \label{eq:B-C-rel3}
\end{eqnarray} 
\par
%
%
Note also that the ladder operators $\hhh^{\pm}(\mh)$ and the real constants
$\llh(\mh)$ corresponding to~(\ref{eq:C-eq}) are given by
\begin{eqnarray}
  \hhh^{\pm}(\mh) & = & \pm \frac{d}{dy} + \frac{\mh+\ch}{y} + \frac{1}{2} \bh y
        \label{eq:C-ladder} \\
  \llh(\mh) & = & - 2\bh \mh + \frac{1}{2} \bh. \label{eq:C-L}
\end{eqnarray}
\par
%
%
In solving equations (\ref{eq:B-C-rel1}) -- (\ref{eq:B-C-rel3}), there are in
principle two indeterminate signs, namely those of $\bh$ and
$\mh+\ch+\frac{1}{2}$. From~(\ref{eq:C-L}), it results that the former determines
whether the type C problem is of class I or II. For definiteness sake, we shall
assume here that the former alternative holds true, so that $\bh \in \R^-$, $\mh =
0$, 1, \ldots,~$\lh$, and
\begin{equation}
  \lambdah = \llh(\lh+1) = - \bh \left(2\lh + \case{3}{2}\right). \label{eq:C-lambda}
\end{equation}
The other case can be treated in a similar way.\par
%
%
Equations (\ref{eq:B-lambda}), (\ref{eq:B-C-rel1}) -- (\ref{eq:B-C-rel3}),
and~(\ref{eq:C-lambda}) then lead to
\begin{equation}
  \mh + \ch + \frac{1}{2} = 2 \epsilon (\lb+\cb) \qquad \bh = - \frac{\db}{\ab} \qquad
  \lh + \ch + \frac{1}{2} = \mb + \cb + \epsilon (\lb + \cb)
\end{equation}
where $\epsilon = \pm 1$. Note that both signs of $\mh + \ch + \frac{1}{2}$ are
allowed due to the finite range of values of $\mh$ characteristic of class I
problems.\par
%
%
It now remains to combine (\ref{eq:F-B-change}) with~(\ref{eq:B-C-change}) to
transform the type F factorizable equation~(\ref{eq:F-eq}) into a type C one. The
result reads
\begin{equation}
  \left[\frac{d^2}{dy^2} - \frac{(2m + \frac{1}{2}) (2m + \frac{3}{2})}{y^2} -
  \frac{q^2}{4(l+1)^2} y^2 - 2q\right] \psih(y) = 0
\end{equation}
the relations between parameters of both factorization types being
\begin{equation}
  \bh = \frac{q}{l+1} = - \sqrt{-\lambda} \qquad \mh+\ch = \epsilon (2m+1) - 
  \frac{1}{2} \qquad \lh+\ch = l + \epsilon \left(m + \frac{1}{2}\right).
  \label{eq:F-C-solrel}
\end{equation}
\par
%
%
In such a process, the type C ladder operators~(\ref{eq:C-ladder}) become ladder
operators for the original type F eigenfunctions $\psi$,
\begin{equation}
  \hhc^{\pm}(l,m) \equiv e^{\ab x/2} y^{-1/2} \hhh^{\pm}(\mh) y^{1/2} e^{-\ab x/2}
\end{equation}
or more explicitly
\begin{eqnarray}
  \hhc^{\pm}_1(l,m) & = & \sqrt{r} \left(\pm \frac{d}{dr} + \frac{2m + \frac{1}{2}
        \mp \frac{1}{2}}{2r} - \sqrt{-\lambda}\right) \label{eq:F-ladder1} \\
  \hhc^{\pm}_2(l,m) & = & \sqrt{r} \left(\pm \frac{d}{dr} - \frac{2m + \frac{3}{2}
        \pm \frac{1}{2}}{2r} - \sqrt{-\lambda}\right) \label{eq:F-ladder2}
\end{eqnarray}
according to whether we choose $\epsilon = +1$ or $\epsilon = -1$. Since the
operators~(\ref{eq:C-ladder}) leave $\lh$ fixed while changing $\mh$ into $\mh
\mp 1$, it follows from~(\ref{eq:F-C-solrel}) that the transformed
operators~(\ref{eq:F-ladder1}) and~(\ref{eq:F-ladder2}) change both $l$ and $m$
into $l \mp \frac{1}{2}$, $m \pm \frac{1}{2}$ and $l \pm \frac{1}{2}$, $m \pm
\frac{1}{2}$, respectively.\footnote{Noninteger values of $l$ and $m$, forbidden in
the original formulation of the factorization method~\cite{infeld}, are allowed in
its generalization by Cari\~ nena and Ramos~\cite{carinena}. The occurrence of
half-integers here is compatible with this extended theory.}\par
%
%
This time Lie algebras can be obtained by introducing two auxiliary variables
$\alpha$, $\beta \in [0, 2\pi)$ and extended eigenfunctions
\begin{equation}
  \Psi_{\mu,\nu}(r,\alpha,\beta) = (2\pi)^{-1} e^{{\rm i} (\mu\alpha + \nu\beta)}
  \psi(r) \label{eq:F-C-extended}
\end{equation}
where
\begin{equation}
  \mu \equiv l - m \qquad \nu \equiv l + m + 1. \label{eq:mu-nu}
\end{equation}
By replacing $\mu$ and $\nu$ by $- {\rm i} \partial/\partial \alpha$ and $- {\rm i}
\partial/\partial \beta$, respectively, and by making the transformations
\begin{eqnarray}
  (2 \sqrt{-\lambda})^{-1/2} e^{\pm {\rm i} \alpha} \hhc^{\pm}_1\left(l -
         \case{1}{4} \pm \case{1}{4}, m + \case{1}{4} \mp \case{1}{4}\right) & \to &
         A_{\pm} \\
  (2 \sqrt{-\lambda})^{-1/2} e^{\pm {\rm i} \beta} \hhc^{\pm}_2\left(l -
         \case{1}{4} \pm \case{1}{4}, m - \case{1}{4} \pm \case{1}{4}\right) & \to &
         B_{\pm}  \label{eq:B-op}
\end{eqnarray} 
we obtain
\begin{eqnarray}
  A_{\pm} & = & \frac{1}{\sqrt{2 \sqrt{-\lambda}}} e^{\pm {\rm i} \alpha} \sqrt{r}
        \left[\pm \frac{\partial}{\partial r} + \frac{1}{2r} \left({\rm i}
        \frac{\partial}{\partial \alpha} - {\rm i} \frac{\partial}{\partial \beta} \mp 1
        \right) - \sqrt{-\lambda}\right] \label{eq:Apm} \\
  B_{\pm} & = & \frac{1}{\sqrt{2 \sqrt{-\lambda}}} e^{\pm {\rm i} \beta} \sqrt{r}
        \left[\pm \frac{\partial}{\partial r} - \frac{1}{2r} \left({\rm i}
        \frac{\partial}{\partial \alpha} - {\rm i} \frac{\partial}{\partial \beta} \pm 1
        \right) - \sqrt{-\lambda}\right] \label{eq:Bpm}
\end{eqnarray}
where we note that $A_{\pm}$ and $B_{\pm}$ only differ by the substitutions
$\alpha \leftrightarrow \beta$, $\partial/\partial\alpha \leftrightarrow
\partial/\partial\beta$.\par
%
%
Contrary to what happens either for type E factorization or for type F factorization
when starting from type B one, the operators $A_{\pm}$ and $B_{\pm}$ do not
belong to su(1,1) algebras, but instead close two commuting Heisenberg-Weyl
algebras:
\begin{equation}
  [A_-, A_+] = [B_-, B_+] = I \qquad [A_{\pm}, B_{\pm}] = [A_{\pm}, B_{\mp}] = 0.
\end{equation}
\par
%
%
{}From (\ref{eq:F-C-extended}) -- (\ref{eq:B-op}), it is obvious that $A_{\pm}$
(resp.\ $B_{\pm}$) change $\mu$ into $\mu \pm 1$ (resp.\ $\nu$ into $\nu \pm 1$),
while leaving $\nu$ (resp.\ $\mu$) and $\lambda$ unchanged. As a consequence, the
parameter $q$ is changed into $q' = q (l+1\pm\frac{1}{2})/(l+1) = q (\mu+\nu+1
\pm1)/(\mu+\nu+1)$.\par
%
%
The interpretation of the $\mbox{\rm w(1)} \oplus \mbox{\rm w(1)}$ algebra
obtained in the present section will become clearer when illustrated on the
example of the Coulomb problem in the next section.\par
%
%
\section{The Coulomb problem}

\setcounter{equation}{0}
In units wherein $\hbar = \mu = e = 1$, the radial wavefunction for an electron in a
Coulomb potential satisfies the equation
\begin{equation}
  \left(\frac{d^2}{dr^2} + \frac{2}{r} \frac{d}{dr} - \frac{L(L+1)}{r^2} + \frac{2Z}{r}
  + 2E\right) R(r) = 0 \label{eq:Coulomb-eq}
\end{equation}
where $L$ is the orbital angular momentum and $Z$ the atomic number. The
negative-energy eigenvalues and corresponding eigenfunctions are given
by~\cite{schiff}
\begin{equation}
  E_n = - \frac{Z^2}{2n^2} \qquad n = n_r + L + 1 \label{eq:Coulomb-E}
\end{equation}
and 
\begin{equation}
  R_{nL}(r) = N_{nL} e^{- \frac{1}{2} \rho} \rho^L L^{(2L+1)}_{n-L-1}(\rho) \qquad
  \rho \equiv \gamma r \qquad \gamma \equiv \frac{2Z}{n} \label{eq:Coulomb-R}
\end{equation}
where $n_r$, $L=0$, 1, 2,~\ldots, $N_{nL}$ is some normalization coefficient, and
$L^{(\alpha)}_n(x)$ is a generalized Laguerre polynomial.\footnote{Contrary to what
is done in~\cite{schiff}, we use here the conventional
definition~\cite{abramowitz} of generalized Laguerre polynomials.}\par
%
%
By setting $R(r) = S(r)/r$, equation~(\ref{eq:Coulomb-eq}) can be rewritten in a
form similar to~(\ref{eq:F-eq}) with
\begin{equation}
  q = - Z \qquad m = L \qquad \lambda = 2E \qquad \psi(r) = S(r).
  \label{eq:F-Coulomb}
\end{equation}
Here $q \in \R^-$ as assumed in sections~2 and~3. Comparing the expressions of
$\lambda$ resulting from~(\ref{eq:F-lambda}) and from~(\ref{eq:Coulomb-E})
and~(\ref{eq:F-Coulomb}), we obtain the relation
\begin{equation}
  l = n-1 = n_r+L \label{eq:l-n}
\end{equation}
between the eigenvalue labels $n$ and $l$, coming from the resolution of the
Schr\"odinger equation and the factorization method, respectively.\par
%
%
Considering first the mapping of~(\ref{eq:Coulomb-eq}) onto a type B factorizable
equation, we get from~(\ref{eq:t}) and~(\ref{eq:l-n})
\begin{equation}
  t = n. \label{eq:t-n}
\end{equation}
By using (\ref{eq:Coulomb-R}), (\ref{eq:F-Coulomb}), (\ref{eq:t-n}), and the results
of~\cite{schiff} and~\cite{abramowitz}, the corresponding extended
eigenfunctions~(\ref{eq:F-B-extended}) can be written as
\begin{equation}
  \Psi^{(m)}_t(r,\eta) = (2\pi)^{-1/2} N^{(m)}_t e^{{\rm i}t\eta} e^{-\frac{1}{2}\rho}
  \rho^{m+1} L^{(2m+1)}_{t-m-1}(\rho) \qquad \rho = \gamma r
  \label{eq:Coulomb-extended}
\end{equation}
where
\begin{equation}
  N^{(m)}_t = \left(\frac{\gamma (t-m-1)!}{2t (t+m)!}\right)^{1/2}.
\end{equation}
When acting on such extended eigenfunctions, the su(1,1) generators $T_{\pm}$ of
equation~(\ref{eq:Tpm}) become
\begin{equation}
  T_{\pm} = e^{\pm{\rm i}\eta} \left(\mp \rho \frac{\partial}{\partial\rho} + {\rm i}
  \frac{\partial}{\partial\eta} + \frac{1}{2} \rho\right).
\end{equation}
\par
%
%
After some calculations using well-known properties of generalized Laguerre
polynomials~\cite{abramowitz}, we obtain
\begin{equation}
  T_{\pm} \Psi^{(m)}_t = - \left(\frac{(t\pm1) (t\mp m) (t\pm m\pm1)}{t}\right)
  ^{1/2} \Psi^{(m)}_{t\pm1}
\end{equation}
which, together with~(\ref{eq:T0-action}), give the action of the su(1,1)
generators in the Coulomb case. The conserved quantity is the angular momentum
$L$. The operators $T_{\pm}$ change $n$ into $n\pm1$ (or $n_r$ into $n_r\pm1$)
and $Z$ into $Z' = Z (n\pm1)/n$, thus leaving the energy unchanged. Such transitions
are relevant to the theory of hydrogen-like ions whenever the ratio $Z/n$ is
integer. As far as we know, this su(1,1) potential algebra for the Coulomb problem
has been noted nowhere else, although it can be easily seen to be equivalent to the
supersymmetric analysis of Haymaker and Rau~\cite{haymaker}.\par
%
%
Considering next the mapping of~(\ref{eq:Coulomb-eq}) onto a type C factorizable
equation, we get from~(\ref{eq:mu-nu}), (\ref{eq:F-Coulomb}), and~(\ref{eq:l-n})
\begin{equation}
  \mu = n-L-1 = n_r \qquad \nu = n+L = n_r+2L+1.
\end{equation}
The extended eigenfunctions are not given by~(\ref{eq:Coulomb-extended})
anymore, but instead by
\begin{equation}
  \Psi_{\mu,\nu}(r,\alpha,\beta) = (2\pi)^{-1} N_{\mu,\nu} e^{{\rm i} (\mu\alpha +
  \nu\beta)} e^{- \frac{1}{2} \rho} \rho^{(\nu-\mu+1)/2} L^{(\nu-\mu)}_{\mu}(\rho)
  \qquad \rho = \gamma r 
\end{equation}
where
\begin{equation}
  N_{\mu,\nu} = \left(\frac{\gamma \mu!}{(\mu+\nu+1) \nu!}\right)^{1/2}.
\end{equation}
When acting on such functions, the w(1) generators $A_{\pm}$ of
equation~(\ref{eq:Apm}) become
\begin{equation}
  A_{\pm} = e^{\pm{\rm i}\alpha} \sqrt{\rho} \left[\pm \frac{\partial}{\partial
  \rho} + \frac{1}{2\rho} \left({\rm i} \frac{\partial}{\partial\alpha} - {\rm i}
  \frac{\partial}{\partial\beta} \mp 1\right) - \frac{1}{2}\right]
\end{equation}
and similarly for the generators $B_{\pm}$ of equation~(\ref{eq:Bpm}).\par
%
%
It is now straightforward to show that
\begin{eqnarray}
  A_{\pm} \Psi_{\mu,\nu} & = & \left(\frac{(\mu+\nu+1\pm1) (\mu + \frac{1}{2}
        \pm \frac{1}{2})}{\mu+\nu+1}\right)^{1/2} \Psi_{\mu\pm1,\nu} \\
  B_{\pm} \Psi_{\mu,\nu} & = & - \left(\frac{(\mu+\nu+1\pm1) (\nu + \frac{1}{2}
        \pm \frac{1}{2})}{\mu+\nu+1}\right)^{1/2} \Psi_{\mu,\nu\pm1}.
\end{eqnarray}
The operators $A_{\pm}$ (resp.\ $B_{\pm}$) change $n$, $L$, and $Z$ into $n \pm
\frac{1}{2}$, $L \mp \frac{1}{2}$ (resp.\ $L\pm \frac{1}{2}$), and $Z' = Z (n\pm
\frac{1}{2})/n$, while leaving the energy unchanged. It should be stressed that the
resulting extended eigenfunctions do not correspond to eigenfunctions of any
physical Coulomb problem since for the latter $n$ and $L$ are restricted to integer
values.\par
%
%
The usefulness of $A_{\pm}$ and $B_{\pm}$ appears when considering their
bilinear products, whose action gives rise to physical eigenfunctions, since the
integer character of all quantum numbers is then retrieved (provided $Z/n$ is
integer). Such bilinear products  generate an sp(4, \R) Lie algebra, which has
various physically-relevant  su(1,1) or u(2) subalgebras. This type of algebraic
description of the Coulomb problem could alternatively be derived from a similar
description of the radial oscillator problem~\cite{fernandez} and the known
mapping of the eigenstates of the Coulomb problem onto the even angular
momentum eigenstates of the four-dimensional oscillator~\cite{nouri}.\par
%
%
\section{Conclusion}

In the present paper, we have completed the analysis of the possibilities
of connections among different factorization types that we had started in I.
It is worth stressing that although some of them have been suggested by
Infeld and Hull~\cite{infeld} and analyzed to a certain extent by other
authors~\cite{gango}, for the first time we have explored them in a general and systematic
way, while providing an algebraization of ladder operators.\par
%
%
More specifically, we have shown that the approach followed in~I and in the
present comment leads to a unified Lie algebraic description of type E and F
factorizable Hamiltonians. The main conclusion of such an analysis is that the existence of 
satellite algebras is a characteristic property of type E factorizable problems. A
similar construction procedure applied to type F ones indeed leads to
operators that do not change the energy eigenvalue and which therefore
generate other types of physically relevant algebras.\par
%
%
This has been illustrated with a detailed discussion of the Coulomb problem,
where we obtained the explicit action of the algebra generators resulting
from the connection of type F factorization with either type B or type C
ones.\par
%
%
As a final point, it is worth mentioning that type D factorization has not been used in our
construction of algebras. Its precise relationship with other factorization types indeed
remains an unsettled issue (see also~\cite{gango}).\par
%
%

\end{document}